\documentclass{aa}  
\usepackage{graphicx}
\usepackage{txfonts}
\usepackage{amssymb}
\usepackage{natbib}
\newcommand{\fun}{\hbox{\ erg cm$^{-2}$ s$^{-1}$} }
\newcommand{\lun}{\hbox{\ erg s$^{-1}$} }
\def\Chandra{{\it Chandra~}}
\def\XMM{{\it XMM-Newton~}}

\begin{document}

\title{Multi-wavelength study of XMMU J2235.3-2557: the most massive galaxy
  cluster at $z>1$\thanks{Based on observations carried out using the
    Advanced Camera for Surveys at the Hubble Space Telescope under
    Program ID 10698; the Very Large Telescope at the ESO Paranal
    Observatory under Program IDs 274.A-5024(B), 077.A-0177(A,B),
    074.A-0023(A), 077.A-0110(A,B)}  }

  \author{P. Rosati,\inst{1}
          P. Tozzi,\inst{2,13}
	  R. Gobat,\inst{3}
          J.S.Santos, \inst{2}
          M. Nonino, \inst{2}
          R. Demarco, \inst{5}
	  C. Lidman,\inst{4,14}
 	  C.R. Mullis, \inst{7} 
          V. Strazzullo, \inst{6}
          H. B\"ohringer,\inst{9}
	  R. Fassbender,\inst{9}
          K. Dawson,\inst{8}
          M. Tanaka, \inst{1}
          J. Jee, \inst{12}
          H. Ford, \inst{11}
	  G. Lamer,\inst{10}
	  A. Schwope\inst{10}         
	}

\institute{
  \inst{1} European Southern Observatory, Karl Schwarzschild Strasse 2, Garching bei Muenchen, D-85748, Germany \\\email{prosati@eso.org}\\
  \inst{2} INAF-Osservatorio Astronomico di Trieste, Via Tiepolo 11, 34131 Trieste, Italy\\
 \inst{3} CEA, Laboratoire AIM-CNRS-Universit\'e Paris Diderot, Irfu/SAp, Orme des Merisiers, F-91191 Gif-sur-Yvette, France \\
  \inst{4} European Southern Observatory, Alonso de Cordova 3107, Casilla 19001, Santiago, Chile\\
 \inst{5} Department of Astronomy, Universidad de Concepci\'on. Casilla 160-C, Concepci\'on, Chile\\
 \inst{6} National Radio Astronomy Observatory, PO box O, Socorro, NM 87801, USA\\
 \inst{7} Wachovia Corporation, NC6740, 100 N. Main Street, Winston-Salem, NC 27101 \\
  \inst{8} Department of Physics and Astronomy, University of Utah, Salt Lake City, UT 84112\\
 \inst{9} Max-Planck-Institut f\"ur extraterrestrische Physik,
              Giessenbachstra\ss e, 85748 Garching, Germany\\
 \inst{10} Astrophysikalisches Institut Potsdam (AIP), An der Sternwarte 16, D-14482 Potsdam, Germany \\
 \inst{11} Department of Physics and Astronomy, Johns Hopkins University, Baltimore, MD21218, USA\\
 \inst{12} Department of Physics, University of California, Davis, One Shields Avenue, Davis, CA 95616, USA\\
\inst{13} INFN, National Institute for Nuclear Physics, Trieste, Italy\\
\inst{14} The Oskar Klein Centre, Stockholm University, S--106~91 Stockholm, Sweden
}

\date{Received 10 August 2009 / Accepted 12 October 2009}

  \abstract
  {The galaxy cluster XMMU~J2235.3$-$2557 (hereafter XMM2235), spectroscopically
    confirmed at $z=1.39$, is one of the most distant X-ray selected
    galaxy clusters. It has been at the center of a multi-wavelength
    observing campaign with ground and space facilities.}
   {We characterize the galaxy populations of passive members, the
     thermodynamical properties and metal abundance of the hot gas,
     and the total mass of the system using imaging data with HST/ACS
     ($i_{775}$ and $z_{850}$ bands) and VLT/ISAAC (J and K$_S$
     bands), extensive spectroscopic data obtained with VLT/FORS2, 
     and deep (196\,ks) \Chandra observations.
       }
   {\Chandra data allow temperature and metallicity to be measured
     with good accuracy and the X-ray surface brightness profile to be
     traced out to 1\arcmin\ (or 500 kpc), thus allowing the mass to be
     reliably estimated.  Out of a total
     sample of 34 spectroscopically confirmed cluster members, we
     selected 16 passive galaxies (without detectable [OII]) within the central
     2\arcmin\ (or 1 Mpc) with ACS coverage, and inferred star formation
     histories for subsamples of galaxies inside and outside the core by
     modeling their spectro-photometric data with spectral synthesis
     models.}
   {\Chandra data show a regular elongated morphology, closely
     resembling the distribution of core galaxies, with a significant
     cool core. We measure a global X-ray temperature of $kT =
     8.6_{-1.2}^{+1.3}$ keV (68\% confidence), which we find to be robust
     against several systematics involved in the X-ray spectral
     analysis. By detecting the rest frame 6.7 keV Iron K line in the \Chandra
     spectrum, we measure a metallicity $Z= 0.26^{+0.20}_{-0.16} \,
     Z_\odot$.  In the likely hypothesis of hydrostatic equilibrium,
     we obtain a total mass of $M_{\rm tot}(<1\,{\rm Mpc})= (5.9\pm 1.3)\times
     10^{14}\,M_\odot$.  By modeling both the composite spectral energy
     distributions and spectra of the passive galaxies in and outside
     the core, we find a strong mean age radial gradient.  Core
     galaxies, with stellar masses in excess of $10^{11} M_\odot$,
     appear to have formed at an earlier epoch with a relatively short
     star formation phase ($z=5-6$), whereas passive galaxies outside
     the core show spectral signatures suggesting a prolonged star
     formation phase to redshifts as low as $z\approx 2$. }
   {Overall, our analysis implies that XMM2235 is the hottest and
     most massive bona--fide cluster discovered to date at
     $z>1$, with a baryonic content, both its galaxy population and
     intracluster gas, in a significantly advanced evolutionary stage
     at 1/3 of the current age of the Universe.
}


\titlerunning{Multi-wavelength study of XMMU J2235.3-2557}
\authorrunning{P.Rosati et al.}

\maketitle

\section{Introduction}

Over the past two decades, considerable effort has been devoted to
discovering ever more distant galaxy clusters using different
observational methods \citep[e.g.][for a
review]{2002ARA&A..40..539R}. These studies have been traditionally
motivated by cosmological applications of the cluster abundance at
high redshift \citep[e.g.][for a review]{2005RvMP...77..207V}, and
also by the use of clusters as laboratories to investigate galaxy
evolution. Clusters provide a convenient and efficient way of studying
large populations of early-type galaxies, which provide stringent
tests on galaxy evolution models in the current hierarchical formation
paradigm \citep{2006ARA&A..44..141R}, because they are the most
massive galaxies with the oldest stellar populations (at least out to
$z\sim\!  2$). Clearly, the higher the redshift the stronger the
leverage on theoretical models. In addition, galaxy properties in
clusters can be contrasted with those in field surveys, which have
multiplied in recent years, thus extending the baseline over which
environmental effects can be studied.

X-ray selection of clusters has been central in these studies, as it
naturally provides gravitationally bound systems (as opposed to simple
overdensities of galaxies) with a relatively simple selection
function. Using ROSAT serendipitous surveys, supported by near IR
imaging and spectroscopy with 8-10m class telescopes, the redshift
envelope was pushed to $z=1.3$, with only 5 clusters discovered at
$z>1$, approximately one per square degree
\citep{2002ARA&A..40..539R}.  The extension of the same technique to
\XMM serendipitous surveys has led to the discovery of two clusters
at $z>1.3$ to date, XMMU J2235.3$-$2557 at $z=1.39$
\citep{2005ApJ...623L..85M} (hereafter M05) and XMMXCS~J2215.9$-$1738
at $z=1.46$ \citep{2006ApJ...646L..13S}. See also
\citet{2008A&A...487L..33L} for a newly discovered, high X-ray
luminosity massive cluster at $z\sim\! 1$.

With the advent of the {\it Spitzer} observatory, an alternative and
efficient way to unveil distant clusters over large areas (as red
galaxy overdensities in the IRAC and optical bands) has been
developed. This technique has given notable results, with three
clusters spectroscopically confirmed at $z>1.3$ in the IRAC Shallow
Survey \citep[ISCS;][]{2008ApJ...684..905E} and one in the SpARCS
survey \citep{2009ApJ...698.1943W}.

Detailed investigations of galaxy populations in the X-ray
luminous clusters at $z>1$ have been conducted with HST/ACS in
combination with the VLT and the Keck telescopes. The study of clusters at
$z=1.10,\, 1.24,\, 1.26,\, 1.27$ \citep{2006ApJ...644..759M,
  2007ApJ...663..164D, 2009ApJ...690...42M} and the aforementioned
systems at $z=1.39$ \citep{2008A&A...489..981L} and $z=1.46$
\citep{2009ApJ...697..436H}, have revealed tight red sequences for the
early-type galaxies, with scatters only marginally larger than those
in local clusters, implying that most of their stellar mass was
assembled at $z>3$ with passive evolution thereafter.  While a change
in the morphology-density relation of early type galaxies (E+S0
galaxies) has been observed at $z\sim\! 1$, elliptical galaxies
still dominate the cluster galaxy population up to $z\sim\! 1.2$
\citep{2005ApJ...623..721P,2007ApJ...670..190H}.

A comparison of cluster and field early-types of similar stellar mass
has revealed a mild but significant difference between the star
formation histories in the two environments
\citep[e.g.][]{2008A&A...488..853G,2008arXiv0806.4604R,2007ApJ...655...30V},
a result which is predicted by current hierarchical galaxy formation
models \cite[e.g.][]{2008ApJ...685..863M}. To date, such comparative
studies can only be carried out at $z\lesssim 1.4$, whereas the
existence of a substantial population of old, massive, passively
evolving early-type galaxies in the field is now well established up
to $z\sim 2$ \citep[][]{2006ApJ...649L..71K,2008A&A...482...21C}.

By pushing cluster studies to higher redshifts, where evolutionary
time scales become comparable to the age of the Universe at these
redshifts, one would expect to detect significant evolutionary
effects.  However, as discussed in this paper, this has not been the
case so far, even after probing two-thirds of the look-back time, not only
for the galaxy populations but also for the thermodynamical properties
and chemical enrichment of the hot gas measured with follow-up
\Chandra observations in $z>1$ clusters.

Our current understanding is that relations, such as the red sequence
in the color-magnitude diagram, the morphology-density relation, the
$L_X-T_X$ relation for the intracluster gas, emerge at $z\lesssim
2$. For example, the study of a proto-cluster at $z=2.16$ identified
around a powerful radio galaxy provided evidence of a forming red
sequence \citep{2008ApJ...680..224Z,2007MNRAS.377.1717K}, which likely
takes 1-2 Gyrs to form \citep{2008A&A...488..853G}.  It is unfortunate
that such a transition in the assembly process of galaxy clusters
seems to occur in a redshift range where spectroscopic observation are
particularly difficult.
In this spirit, we have carried out a multi-wavelength study of one
the most distant clusters known, XMMU J2235.3$-$2557 (hereafter
XMM2235) at $z=1,39$, which was the first distant cluster confirmed
(M05) as part of the on-going \XMM Distant Cluster Project
\citep[XDCP,][]{2005Msngr.120...33B,2008arXiv0806.0861F}.  In this 
paper, we use spectro-photometric observations of XMM2235 in the
optical/near-IR and X-ray bands to characterize its galaxy population
(particularly the passive spectroscopic members) and the thermodynamic
status of the hot intracluster gas and to measure its total mass. In
section \ref{sec:VLTHST}, we present the VLT and HST data used in this
paper, including an extensive spectroscopic campaign which yielded 34
confirmed cluster members. In section \ref{sec:SFH}, we model the
underlying stellar populations of passive galaxies and constrain their
star formation histories. In section \ref{sec:Xray}, we present deep
\Chandra observations of XMM2235, the methods of analysis and the
resulting measurements of its temperature, metallicity and mass. In
section \ref{sec:results}, we discuss the results.

$H_0=70\ {\rm km}\ {\rm s}^{-1}\,{\rm Mpc}^{-1}, \,\Omega_m=0.3, \,
\Omega_\Lambda=0.7$ are adopted throughout this paper. In this
cosmology, 1\arcmin\ on the sky corresponds to 0.5 Mpc at $z=1.39$

\section{VLT and HST Observations}
\label{sec:VLTHST}

\begin{figure*}
\centering
\includegraphics[width=12cm,angle=0,clip=true]{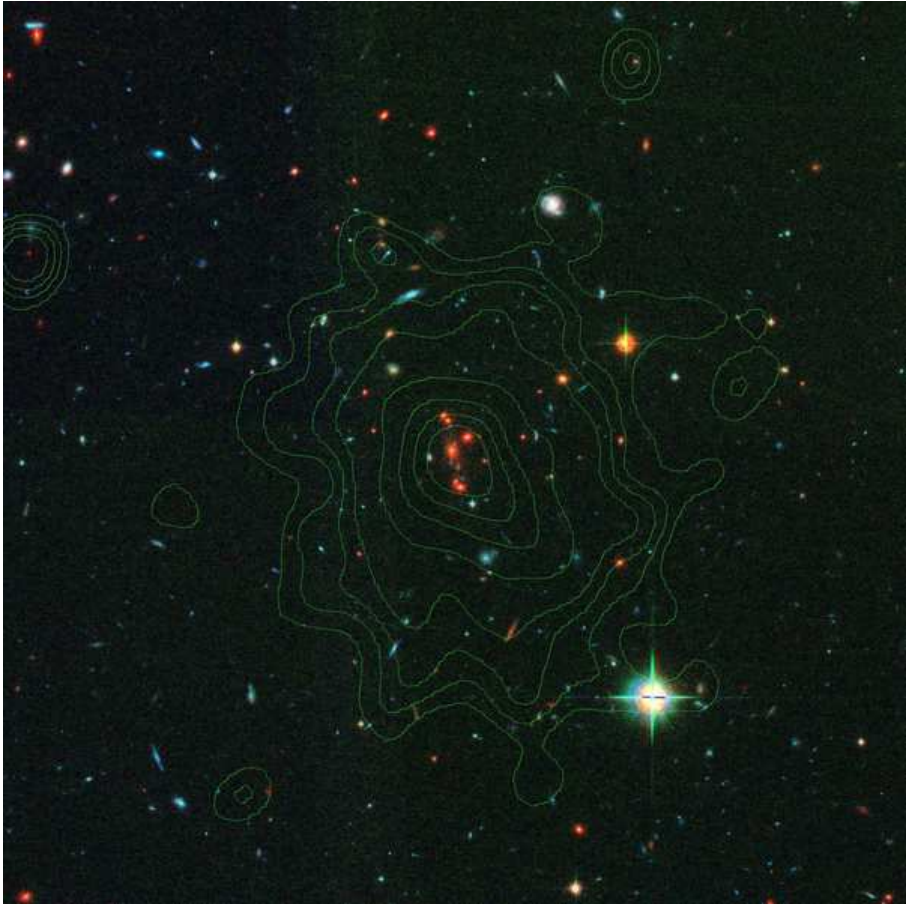} 
\caption{Color image of XMM2235 obtained from the combination of $i$,
  $z$ (HST/ACS) and $K_s$ (VLT/ISAAC) filters. Overlaid X-ray contours
  are from a \Chandra 196 ks observation (smoothed by a gaussian with
  $\sigma=2\arcsec$; the lowest contour corresponds to 3$\sigma$ above the background). The image is 2\arcmin\ across, corresponding to
  $1.0 h_{70}^{-1}$ Mpc at $z=1.4$, and is centered on
  $\rm{RA}=22^h35^m20^s.81$, $\rm{Dec}=-25^\circ57\arcmin40\arcsec$. North is up, east is to the left.}
\label{f:xfield}
\end{figure*}

\subsection{Optical and near IR data}
\label{sec:OptData}

Following the relatively shallow discovery imaging data obtained with
the VLT/FORS2 in the R and z bands (M05), XMM2235 was
observed in the J and K$_S$ bands with ISAAC on the VLT in January
2006 and August 2007.  More recently, deeper and wider-field near IR
observations of XMM2235 were obtained with the HAWK-I instrument on
the VLT and presented in \cite{2008A&A...489..981L}.

The ISAAC observations used here consist of a mosaic of $3\times 3$
pointings, covering $7.5\times 7.5\arcmin$, with exposure times of 30
and 20 minutes each in the J and K$_S$ filters, respectively (program
ID 274.A-5024(B)), with the exception of the central pointing which
was observed for 60 and 45 minutes, respectively (ID
077.A-0177(A)). The ISAAC data were reduced with the ESO/MVM
software\footnote{http://archive.eso.org/cms/eso-data/data-packages/eso-mvm-software-package}. Photometric
zero points (ZP) were derived by observing one or two standards each
night from the photometric catalogue of
\cite{1998AJ....116.2475P}. All images were laid on an astrometric
grid with a pixel scale of 0.15\arcsec that was based on VLT/FORS2
images. In order to derive a uniform ZP across the mosaic, we used
independent observations obtained with the SofI camera on the ESO-NTT,
also reduced with ESO/MVM, for which accurate ZPs were derived with
several photometric standards observed before and after the
observations. These stars yielded rms scatters of $<0.02$ mag in both
J and K$_S$. The $5\times 5\arcmin$ SofI field includes the central
ISAAC field and covers partially the other eight tiles, thus allowing
the photometric uniformity across the mosaic to be controlled and
independently quantified. The FWHM of all stars in each field was
measured, ranging from 0.4 to 0.7\arcsec. Gaussian smoothing was used
to obtain a uniform PSF in J and K$_S$ mosaiced images (with a FWHM of
$\!=0.51\arcsec$ in K$_S$ and $0.73\arcsec$ in J).  A sample of 62
stars across the mosaic were selected, using size information from the
ACS images and their $(J-K_S)_{AB}$ color, which was restricted to
$(J-K_S)_{AB} < 1$, from the SofI images. These stars were used to to
compare photometry across all fields, using SofI ZPs as fiducial
ones. As a result, the ZPs in three ISAAC tiles were adjusted by a few
\% to match the SofI photometry, so as to produce mosaics with a
uniform ZP, whose accuracy was estimated to be 0.03 mag in K and 0.04
mag in J as based on the rms scatter from stellar photometry with
respect to SofI photometry. We also checked that these ZPs are
consistent with 2MASS photometry using a small number of unsaturated
stars in the 2MASS catalog.

XMM2235 was observed with the ACS Wide Field Camera on the Hubble
Space Telescope on June 27 2005 in the F7755W and F850LP passbands
(program ID 10698), as part of the ACS Intermediate Redshift Cluster
Survey \citep{2004ASSL..319..459F}. Subsequent ACS visits of XMM2235
took place in 2006 as part of the HST Cluster Supernova Survey (Dawson
et al. 2009). This work, however, is based on the first visit only,
amounting to an integration of 5060 sec in $i_{775}$ and 6240 sec in
$z_{850}$. Analysis of the full HST data set are presented in
accompanying papers \citep[][Strazzullo et al. in
prep.]{2009ApJ...704..672J}. The ACS data were processed with the
latest version of the ``Apsis'' pipeline described by
\cite{2003ASPC..295..257B}. We used the AB photometric system
calibrated with the ACS/WFC zero points from
\cite{2005PASP..117.1049S}.

The ISAAC mosaic images were aligned with the ACS images and a four
band ($i_{775}$, $z_{850}$, J and K$_S$) photometric catalog was
constructed using SExtractor.  Magnitudes were first measured in
apertures of 0.75\arcsec\ radius in the ACS images and 1\arcsec\ in
the ISAAC images, then corrected to 2 and 4\arcsec, respectively, to
take into account the PSF difference between ACS and ISAAC. Such
aperture corrections were derived from the median growth curve of
unsaturated stars in the field. Magnitudes were also corrected for
galactic extinction according to Schlegel (1998). The adopted
corrections are 0.043 mag in $i_{775}$, 0.032 in $z_{850}$, 0.019 in J
and 0.008 in $K_S$.  A color-composite image of XMMU J2235.3-2257 is
shown in Fig.\ref{f:xfield}, with the Chandra X-ray contours overlaid
in green (see Sect.\ref{sec:Xray}). The peak of the X-ray emission is
within 2\arcsec\ of the brightest cluster galaxy (BCG) and is
elongated, from the southwest to the northeast, following the
distribution of the red cluster galaxies in the core and the major
axis of the BCG.

\subsection{VLT Spectroscopy}
Three separate observing runs were originally devoted to the
spectroscopic follow-up study of XMM2235 using FORS2 on the VLT. Two
runs were carried out using the MXU mode, programme ID 074.A-0023(A)
in October 2004 (see M05) and ID 077.A-0177(B) in July 2006, using two
slit masks each. More spectroscopic observations were carried out in
July 2006, as part of a program to search for distant Type Ia
supernovae hosted by early type galaxies in galaxy clusters at $z>1$
(Dawson et al. 2009).  For the latter, the MOS mode of FORS2, which
consists of 19 movable slits that can be moved into the focal plane,
was used.  This allowed rapid follow-up of transient events, such as
Type Ia Supernova. In the assignment of slits, transients were given
highest priority, likely cluster members were given second priority
and other galaxies were given third priority. Two masks were used to
follow three transients. Total integration times varied from 2800
seconds in the first mask to 5400 seconds in the second. The number of
slits that were used in MXU mode was typically more than double the
number that were used in MOS mode, with greater flexibility in placing
targets on slits.  All observations were carried out with the 300I
grism which has a dispersion of $2.6\AA$/pixel, corresponding to a
resolution of $\sim\!10\AA$ for the adopted slit width of 1\arcsec.
In combination with the OG590 order blocking filter, this
configuration provides a wavelength range of $5900-10000\AA$. In the
077.A-0177(B) run, the OG590 filter was not used thus extending the
coverage down to $\sim\!5500\AA$.  For observations with the MXU
masks, integration times ranged from 4 to 8 hours, depending on the
faintness of the targets in a given mask.

Targets were selected using colors and magnitudes, however not
homogeneously throughout the whole observing campaign. Galaxies lying
on the $R-z$ red sequence, with $z_{AB}\lesssim 23$, were selected
first (M05). Then fainter objects with colors consistent with late and
early type galaxies at the cluster redshift were targeted in
subsequent masks. As more photometric information (ACS-$i,z$,
ISAAC-$J,K_s$) became available after the first run, we adopted new
criteria, using K$_{AB}<23$ and $i-K_s$, $J-K_s$ colors ranging from
the bluest confirmed star-forming member to the reddest member of the
red sequence (with a 0.3 mag conservative margin). This corresponded
to $1.3<(i_{775}-K_s)_{\rm AB}<3.7$.

Spectra were reduced with standard techniques using IRAF tasks (see
e.g.~\cite{2007ApJ...663..164D}). The RVSAO/XCSAO routine was used to
measure redshifts via cross-correlations with a range of template
spectra. Redshifts for faint late type galaxies were often derived
from the [OII] line alone. Based on signal-to-noise and
cross-correlation coefficients (R), spectra were finally assigned
quality flags of A,B,C depending on whether the redshifts was
considered secure ($R>4$), acceptable ($R\approx 3$), and unknown
respectively. The first three observing runs yielded 129 redshifts of
quality A or B, from 168 extracted spectra, in a field covering
$\sim\! 3\times 3\arcmin$ (or $1.5\times 1.5$ Mpc). From this sample,
28 redshifts (24 A's$+$4 B's) lying within $\pm 2000$ km/s of the
median redshift $z_{CL}=1.390$ were selected as cluster members,
twelve of which (11 A's$+$1 B's) are star forming galaxies with a
detectable [OII]$\lambda\lambda$\,3727\,\AA\ line (EW([OII])$< -5\AA$
). Magnitudes of cluster members range from $z_{AB}\simeq 21.6$ (the
BCG) to the spectroscopic limit of $z_{AB}\simeq 24$.

More recently, a fourth spectroscopic programme (ID 081.A-0759D, PI:
M.Tanaka), mostly aimed at the outskirts of the cluster, was carried
out in July-August 2008 with the same 300I grism using two MXU
masks. $J-K_s$ colors were used to target galaxies at the cluster
redshift over an area of 3 Mpc radius. These observations yielded 6
extra passive members, two of which are within the central $2\arcmin$ and
are therefore included in the sample used in this paper.

\begin{figure}
     \centering
    \includegraphics[width=0.9\columnwidth]{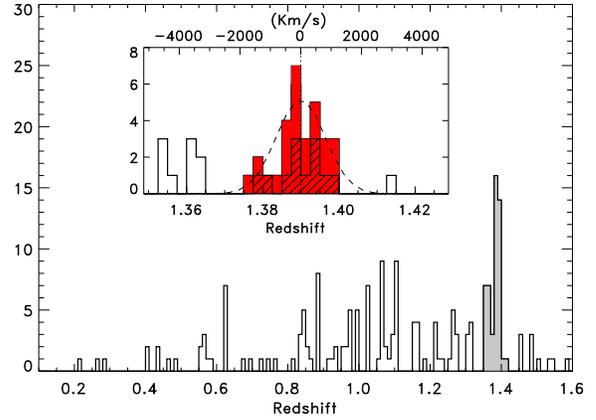}
    \caption{Measured redshifts of galaxies in a field of 3\arcmin\
      around XMM2235 (with a bin of 0.01). The inset shows a blow-up
      of the distribution (grey histogram in the main figure) around
      the cluster redshift, with a bin $\Delta z= 0.0025$. The top
      axis gives the rest frame velocity centered on the median value
      $z=1.390$. The red histogram includes cluster members within
      $\pm 2000$ km/s. The hatched region denotes star-forming
      members. The dashed curves is a gaussian with the estimated
      velocity dispersion of 802 km/s.  }
\label{f:zhisto}
\end{figure}

In Fig.\ref{f:zhisto}, we show the redshift distribution 30 cluster
members, 18 passive and 12 star-forming galaxies. The latter all lie
at projected distances larger than $250$\,kpc. We estimated the
velocity dispersion using the ROSTAT algorithm of
\cite{1990AJ....100...32B}, which yields $\sigma_v=802^{+77}_{-48}$
km/s with formal boostrap errors. A detailed dynamical analysis of the
entire redshift sample will be presented in a forthcoming paper. From
the redshift catalog, we defined a sample of 16 passive galaxies,
which are covered by the ACS and ISAAC data, and whose
spectrophotometric properties are modeled in the following section to
age-date their underlying stellar populations. A complete catalog with
redshifts from the whole spectroscopic campaign, as well as
photometric and morphological measurements using the enhanced HST/ACS
observations, HST/NICMOS and VLT/HAWK-I data, will be presented in a
forthcoming paper (Strazzullo et al., Nunez et al. in prep.). We have
verified that the $i,z,J,K_S$ photometry used in this paper is in very
good agreement with the one in the final augmented imaging data set.

\section{Star-formation histories of passive members}
\label{sec:SFH}

The spectro-photometric data described in the previous section were
used to constrain ages and star formation histories (SFHs) of the 16
passive galaxies (i.e., those with no detectable [OII] emission)
covered by the ACS and ISAAC data.  To this end, we used the technique
described in \cite{2008A&A...488..853G}, which combines both the
spectral energy distributions (SED) and spectra of galaxies to
characterize the underlying stellar populations with spectral
synthesis models. This generally allows stronger constraints to be
obtained, by mitigating inherent degeneracies among stellar population
parameters.  The same technique was used in
\citet{2009A&A...501...49S} to age date early-type galaxies in a
cluster at $z\simeq 1$.  ACS imaging shows that all the galaxies have
early-type morphologies, however we defer the morphological analysis
to another paper (Strazzullo et al. in prep.).

We used a grid of Bruzual\&Charlot (\citeyear{2003MNRAS.344.1000B},
BC03) $\tau$-models, characterized by a delayed exponential
star-formation rate (SFR) $\psi_\tau(t)=(t/\tau^2)\exp(-t/\tau)$,
where $t$ is the time since the onset of star formation and $\tau$ a
characteristic time scale (corresponding to the time of the SFR
peak). Solar metallicity and a Salpeter initial mass function were
adopted. Since spectra extend to rest-frame wavelengths shorter than
3200\AA, we used BC03 templates which include the
\citet{1998PASP..110..863P} stellar library in the mid-UV. SEDs are
constrained by the four photometric bands described in
Section~\ref{sec:OptData}: J, K$_S$, $i_{775}$, $z_{850}$. Stellar
masses for the 16 passive galaxies are obtained, together with ages,
T, and characteristic times, $\tau$, by fitting the SEDs with the
constraint that T is bound to be less than $4.22$ Gyr, which limits
the star formation epoch to $z<15$. Stellar masses range between
$5\times 10^{10}$ and $5\times 10^{11} \rm{M}_\odot$ apart from the
BCG. The four red galaxies in the core (within 100 kpc radius) all
exceed $10^{11}\rm{M}_\odot$ with the BCG being by far the most
massive of the four, with $M_{*BCG}\approx 9\times 10^{11}{M}_\odot$.

\begin{figure}
     \centering
    \includegraphics[width=0.9\columnwidth]{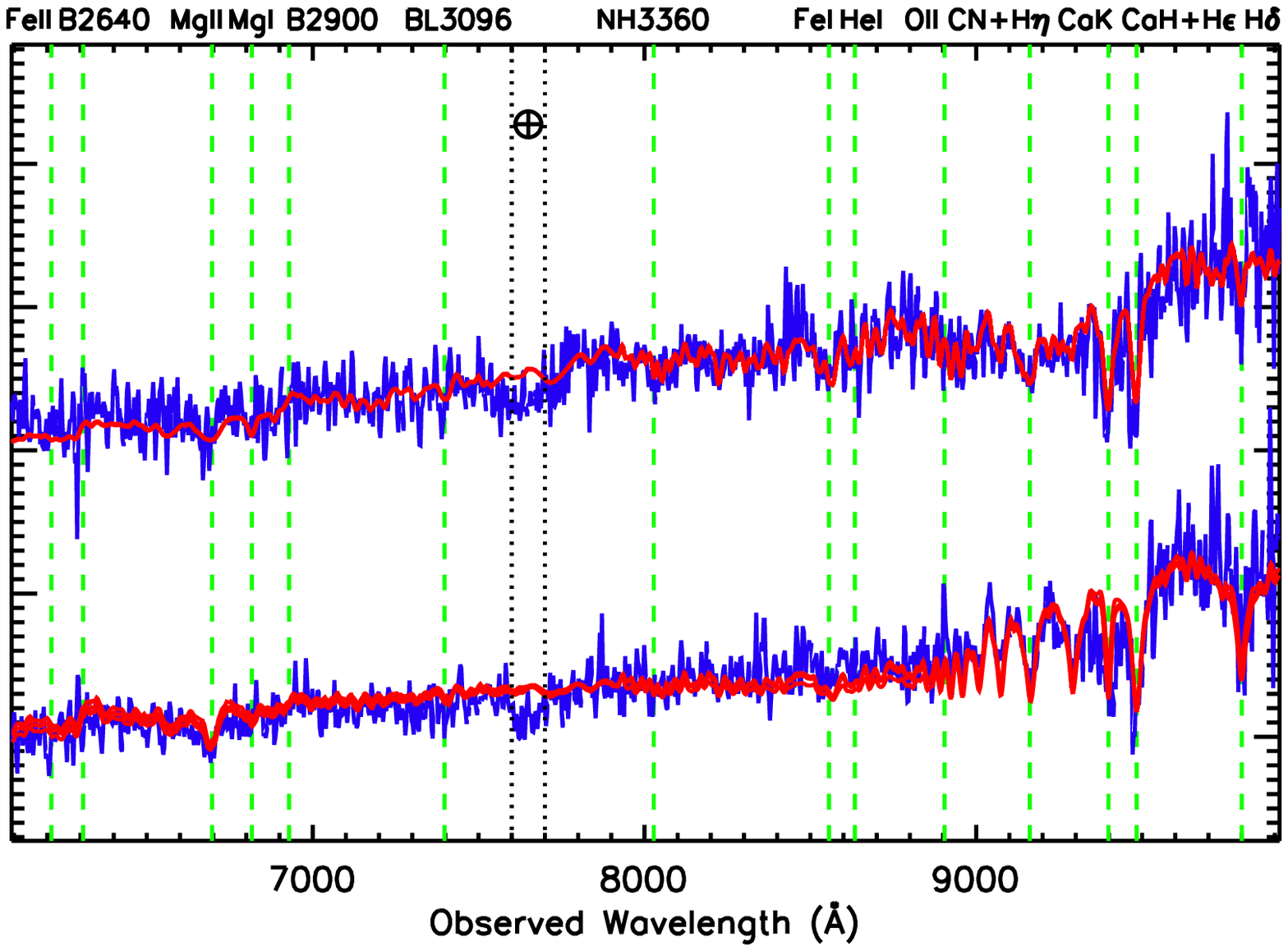}
    \caption{Composite spectra of the four brightest galaxies in the
      core ($R_{CL}<100$\,kpc) of XMM2235 (top) and of the
      spectroscopically passive members (12) in the outskirts
      ($150<R_{CL}<1000$)\,kpc (bottom). Best fit BC03 models to the
      spectro-photometric data of each subsample are overploted in
      red. The $\oplus$ symbol corresponds to the atmospheric A band. This
      region is not used in the analysis.}
\label{f:optspec2}
    \includegraphics[width=0.9\columnwidth]{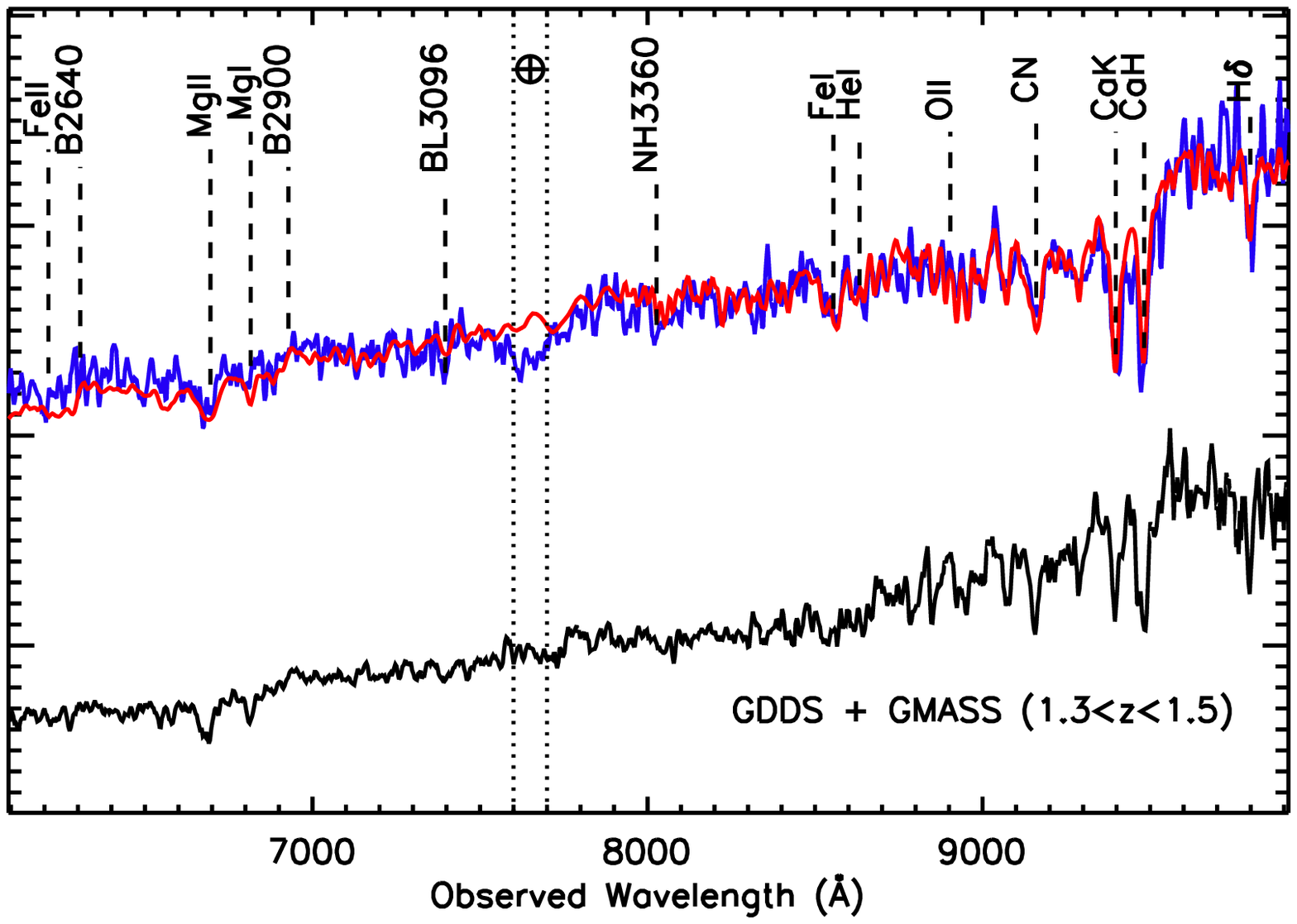}
    \caption{Top: composite spectrum of the 16 confirmed passive
      galaxies of XMM2235 with the best fit BC03 model overplotted in
      red. Bottom: an average spectrum of early-type galaxies at
      a similar redshift from the GDDS and GMASS spectroscopic surveys
      in the field.}
\label{f:optspec}
\end{figure}

\begin{figure}
     \centering
    \includegraphics[width=0.9\columnwidth]{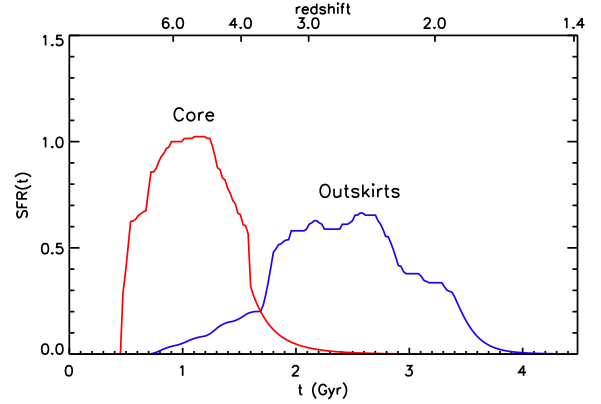}
    \caption{Star formation histories (in arbitrary units)
      derived from the fit of the composite spectro-photometric
        data with BC03 $\tau$-models for the sample of passive
        galaxies in the core and outskirts of
      XMM2235. The lower x-axis shows cosmic time; the upper
      x-axis marks the corresponding redshift.  }
\label{f:SFH}
\end{figure}

In order to have a more robust estimate of the SFHs of massive passive
galaxies in XMM2235, we stacked the SEDs and spectra of all 16 members
and then separately the four in the core and the 12 lying in the
outskirts at cluster-centric distances $150 < R_{CL} < 1000$ kpc
(Figs.\ref{f:optspec2} and \ref{f:optspec}). Note that these coadded
spectra have a similar S/N ratio, despite the difference in the number
of galaxies in each sample. This is due to the higher luminosities of
the core members. The composite spectra have a $S/N\sim\! 7$, making
the fit with BC03 models meaningful.  As described in
\cite{2008A&A...488..853G}, we reconstruct the SFH for a given
subsample by first fitting their average SED, which yields a $1\sigma$
confidence region in the ($T,\, \tau$) space. The subgrid of models
within this region is then used to find the best fit solution to the
average spectrum. Finally, the SFH which best fits the
spectro-photometric data is obtained by computing the mean, at each
epoch, of all models $\psi(T,\tau)$ lying within the 3$\sigma$
confidence region of the $\chi^2$ fit to the average composite
spectrum and average SED (22 models for the outskirts sample and 13
for the core sample). The result is shown in Fig.\ref{f:SFH}. A stark
difference in the derived SFHs of galaxies in the core and the cluster
outskirts is apparent. Galaxies in the core have formed their stars at
early epochs, with star-formation weighted ages of
$T_{SFR}=3.4^{+0.2}_{-0.3}$ Gyr, consistent with a single burst of
star formation at $z_F=5.3^{+1.4}_{-0.8}$. Those lying outside the
core have $T_{SFR}=2.0^{+0.5}_{-0.5}$ Gyr, corresponding to
$z_F=2.7^{+0.6}_{-0.5}$, and a final formation redshift (when 99\% of
the stellar is assembled) of $z_{fin}=2.0^{+0.5}_{-0.2}$.  We note
that the areas of the SFH curves in Fig.\ref{f:SFH} are arbitrarily
normalized to one, since we are only interested in the difference of
the SF time scale between the two samples. It is clear that the
inferred SFH is model dependent and is generally not a unique
solution. The difference in the underlying stellar populations of the
two samples is significant, however, and is evident from the
inspection of their composite spectra and best fit BC03 models
(Fig.\ref{f:optspec2}).  When compared with galaxies in the core,
galaxies in the outskirts show significant post-starforming features
(deep balmer lines) as well as a smaller 4000\AA\ break in addition to
having a bluer average SED. Galaxies in the outskirts show residual
star forming activity about $\sim\! 1$ Gyr before the epoch of
observation. Note that the [OII] line is not detected even in the
stacked spectrum of all the passive galaxies, consistent with no
significant star formation at $z=1.4$.

These findings are in very good agreement with those of
\cite{2008A&A...489..981L}, who used the mean color and scatter of a
high-quality $J-K_S$ color-magnitude diagram of XMM2235, based on
VLT/HAWK-I measurements, to infer a formation redshift of $z_F=4-5$
for galaxies in the cluster core ($R_{CL}<90$ kpc), in contrast with
those outside the core which instead exhibit a larger scatter and
bluer colors. The SFHs in Fig.\ref{f:SFH}, which was derived from
independent spectro-photometric data, elucidate the difference between
these two populations.  This result is consistent with a scenario
where the central galaxies formed rapidly at high redshift and ceased
forming stars early while galaxies outside the immediate core
continued star formation for at least 1 Gyr longer, or alternatively
formed most of their stars $\gtrsim 1$ Gyr later.

The average spectrum of the 16 passive galaxies and the best fit BC03
model to the spectro-photometric data are shown in
Fig.\ref{f:optspec}. For comparison, we also show the composite
spectrum of {\sl field} early-type galaxies at a similar redshift and
covering are similar range of stellar masses, which was obtained by
combining 11 GDSS \citep{2004AJ....127.2455A} and 3 GMASS
\citep{2008A&A...482...21C} spectra in the range $1.3<z<1.5$. It is
interesting to note that, similar to what was found by
\cite{2008A&A...488..853G} \citep[see also][]{2008arXiv0806.4604R},
who compared the SFHs of early-type galaxies in a massive cluster at
$z=1.24$ with those in the field, the 4000\AA\ break in cluster
early-type galaxies appears to be larger than those in field galaxies,
suggesting a higher formation redshift or a shorter duration of the
star forming phase for early-type galaxies in massive clusters.
\cite{2008ApJ...685..863M} found this observational scenario to be in
good agreement with expectations from semi-analytic models of galaxy
formation in which SFHs of cluster galaxies of a given mass are
modulated by environmental effects.

\section{Chandra Observations}
\label{sec:Xray}

XMM2235 was observed with the \Chandra ACIS--S detector in VFAINT mode
in five exposures of 44 ks (Obs ID 6975), 24 ks (Obs ID 6976), 80 ks
(Obs ID 7367), 33 ks (Obs ID 7368) and 15 ks (Obs ID 7404).  The data
were reduced using the Chandra CIAO software V4.1 (December 08
release) and related calibration files, starting from the level 1
event file as described in \citet{2004AJ....127..230R}.  The total
effective exposure time is 196 ks after the application of this
reduction procedure.

In Fig.~\ref{f:xfield}, we show a color composite image combining the
HST/ACS and VLT/ISAAC observations with overlaid \Chandra X-ray
contours in the soft band.  The diffuse X--ray emission from the
cluster can be traced out to $r\simeq 1\arcmin$ ($2\sigma$ above the
background), corresponding to 0.5 Mpc.  The \Chandra image immediately
shows that point sources do not significantly contaminate the cluster
emission, as it is sometimes the case in distant clusters
\citep[e.g.][]{2001ApJ...552..504S, 2008A&A...489..967B}.

\subsection{X--ray Spectral analysis}
We performed a global spectral analysis in an elliptical region with
semi-major and semi-minor axes $a=42\arcsec$ and $b=38\arcsec$
centered on the peak of the photon distribution (RA$=22^h35^m20^s.82$,
Dec=$-25^\circ57\arcmin40.3\arcsec$, J2000), after masking out point
sources.  In this aperture, which maximizes the S/N, we detected
approximately 1900 net counts in the 0.3--10 keV band.  The background
was obtained from a large annulus around the cluster position, after
subtraction of point sources.  We also avoided regions where missing
columns significantly decrease the background intensity.  The
background photon file was scaled to the source file by the ratio of
the geometrical area.  The response matrices and the ancillary
response matrices were computed for each exposure with the tool {\tt
  mkwarf} and {\tt mkacisrmf} applied to the extracted regions.

The spectra were analyzed with XSPEC v12.3.1
\citep{1996ASPC..101...17A} and fitted with a single temperature MEKAL
model (\cite{1995ApJ...438L.115L}), where the ratio between the
elements are fixed to the solar value as in
\citeauthor*{1989GeCoA..53..197A} (1989, hereafter {\sl angr}).  These
values for the solar metallicity have recently been superseded by the
new values of \cite*{1998SSRv...85..161G}, who used a 0.676 times
lower Fe solar abundance \citep[see also][]{2005ASPC..336...25A}.
However, we prefer to report Iron abundance in units of {\sl angr}
since most of the literature still refers to these old values.  Since
our metallicity depends only on the Fe abundance, updated
metallicities can be obtained simply by rescaling by 1/0.676 the
values reported here.  We modeled the Galactic absorption with the
tool {\tt tbabs} \citep*[see][]{2000ApJ...542..914W}.

Spectral fits were performed over the energy range 0.5--7 keV.  We
exclude photons with energy below 0.5 keV in order to avoid systematic
biases in the temperature determination due to uncertainties in the
ACIS calibration at low energies.  We checked that the measured
temperature does not depend significantly on the lower bound, $E_1$,
of the energy range (by changing $E_1$ in the range 0.3--1.2 keV, T
changes by $\sim\! 3\%$).  We also exclude energies above 7 keV where
the signal is dominated by noise.  We used three free parameters in
our spectral fits: temperature, metallicity and normalization.  The
local absorption was frozen to the Galactic neutral hydrogen column
density $N_H = 1.5 \times 10^{20}$ cm$^{-2}$, as obtained from radio
data \citep*{1990ARA&A..28..215D}, and the redshift was set to
$z=1.39$, as measured from the optical spectroscopy.  Spectral fits
were performed using Cash statistics (as implemented in XSPEC) of
source plus background photons, which is preferable for low
signal--to--noise spectra.  We also performed the same fits with
$\chi^2$ statistics (with a standard binning with a minimum of 20
photons per energy channel in the source plus background spectrum) and
verified that our best--fit model gives a reduced $\chi^2 \sim 1.03$
for 107 degrees of freedom. All quoted errors below correspond to
$1\sigma$, or the 68\% confidence level for one interesting parameter.

\begin{figure}
     \centering
    \includegraphics[width=0.9\columnwidth,viewport=0 10 623 510,clip]{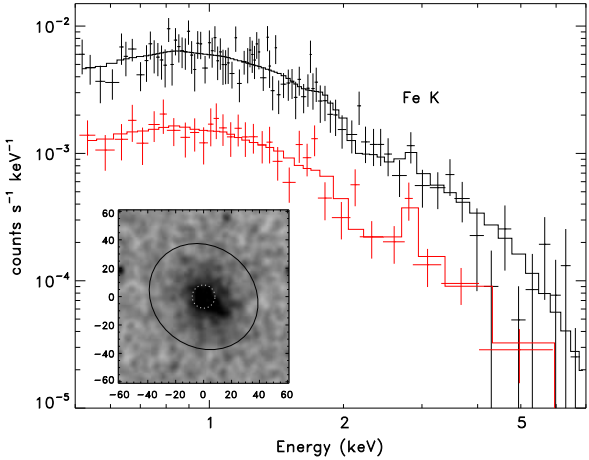}
    \caption{Folded X--ray spectra (data points) and best fit MEKAL
      model (solid line) from \Chandra observations (196 ks) of
      XMM2235 at $z_{\rm spec}=1.39$. The top spectrum is extracted
      from the total elliptical aperture with semi-axis $a=42\arcsec$,
      $b=38\arcsec$; the one on the bottom from the cluster core
      (8\arcsec\ radius). The inset shows a cutout ($120\times
      120\arcsec$) of the cluster in the 0.5-2 keV band (smoothed with
      a $\sigma=2\arcsec$ gaussian) with the two extraction regions.
      A significant redshifted Fe 6.7 keV line is visible.}
\label{f:xspec}
\end{figure}

\begin{figure}
  \centering
  \includegraphics[width=0.9\columnwidth]{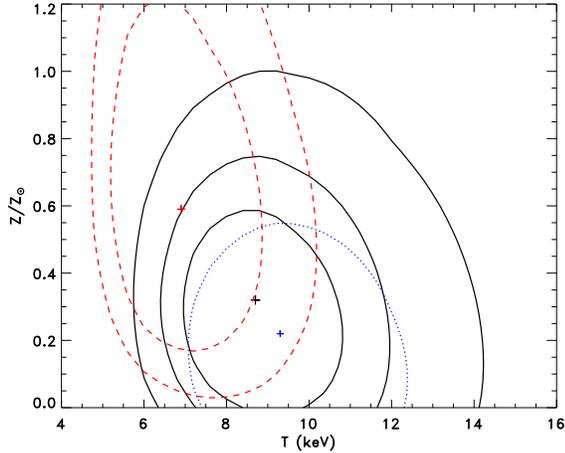}
  \caption{Confidence contours and best fit temperature and iron
    abundance (crosses) of XMM2235 obtained from the spectral fit of
    the \Chandra data (Fig.\ref{f:xspec}). Solid contours ($1, 2,
    3\sigma$ confidence levels for two interesting parameters) refer
    to the total elliptical region, dashed lines ($1, 2\sigma$) to the
    inner core, dotted line ($1\sigma$) to the total region with the
    core removed.  }
\label{f:contours}
\end{figure}

The \Chandra folded spectra of XMM2235, extracted from the whole
elliptical area and the inner core, are shown in
Fig.~\ref{f:xspec}. The corresponding extraction regions are indicated
on the inset image. A significant Fe K line is visible at $E\simeq
2.8$ keV.  Confidence contours from the spectral fit in the $Z-kT$
plane are shown in Fig.~\ref{f:contours}. The fit to the global
spectrum gives a best fit temperature of $kT = 8.6_{-1.2}^{+1.3}$ keV,
and a best fit metallicity of $Z = 0.26^{+0.20}_{-0.16} \, Z_\odot$
($1\sigma$ error bars).  By leaving Galactic absorption as a free
parameter, we obtained $N_H = 1.4_{-0.14}^{+2.2} \times 10^{20}$
cm$^{-2}$, thus consistent with the Galactic value.  In this case, the
best fit temperature is consistent (within $1\sigma$) with the
aforementioned best fit value. Interestingly, leaving the redshift
free, a four parameter fit yields $z = 1.37_{-0.06}^{+0.04} $, which
shows how accurately the redshift can be determined from the X-ray
data alone due to the detection of the iron line.

The best fit model corresponds to an unabsorbed flux within the
elliptical aperture of $(2.7\pm 0.1)\times 10^{-14}\fun $ in the 0.5-2 keV
band.  This corresponds to a luminosity of $(2.1\pm 0.1)\times 10^{44}\lun
h_{70}^{-2}$ in the rest frame 0.5-2 keV band, and a bolometric
luminosity of $(8.5\pm 0.4)\times 10^{44} h_{70}^{-2} \lun$. One can use the
best fit double $\beta$-model of the surface brightness profile
described below to extrapolate these luminosities at larger radii. For
example, values need to be multiplied by a factor 1.3 to encircle the
flux within $r=1 h_{70}^{-1}$ Mpc. Thus, the bolometric luminosity is
$L_{X,bolom}=1.1\times 10^{45} h_{70}^{-2}$\lun. We note that the X-ray
luminosity and temperature of XMM2235 are consistent with the
$L_{X,bolom}-T$ relation determined for hot clusters at $z<1$
\citep[e.g.][]{2002ARA&A..40..539R}.

We performed several checks in order to assess the robustness of our
results against systematics. By changing background regions,
e.g. choosing different external circular regions as opposed to the
annulus above, the best fit temperature varies by only a few \%.  More
recently, a possible calibration problem of ACIS-S at high energies
has been under discussion, which has supposedly been alleviated by the
latest calibration files used here. It has been claimed that it biases the
temperature of hot low-$z$ clusters to higher values, an effect that
is limited to 5\% when the energy range is restricted to 5 keV
(L. David, private communication). We note that by using \Chandra
calibrations of 2007 shortly after the observations, we obtain a
temperature of 9.2 keV (still consistent with the current value at
the $1\sigma$ level), whereas the best fit value remains stable over the
last two CIAO software releases. We also note that the bias is
reduced, because, at $z=1.4$, the exponential Bremsstrahlung cut off
is observed at significantly lower energies. By repeating our fit in
the energy range $0.5-5$ keV, we find that the best fit temperature
decreases by only 0.04 keV.

As shown below, the Chandra data reveal a significant excess in the
center of the X-ray surface brightness (SB) profile of XMM2235, which
is usually interpreted as the presence of a cool core. A spectral
analysis of the inner 8\arcsec (67 kpc) radius, containing only 400
net counts, yields a best--fit temperature $kT = 6.7^{+1.3}_{-1.0}$
keV, and an Iron abundance of $Z = 0.59^{+0.34}_{-0.28} \, Z_\odot$
(dashed contours in Fig\ref{f:contours}). This suggests the presence
of a cool core with higher Iron abundance, a result that needs
significantly more statistics to be confirmed.  We also repeated the
spectral analysis removing this cool core region of 8.0\arcsec radius
(the dotted 1-$\sigma$ contour in Fig.~\ref{f:contours}) from the
total elliptical region and find $kT = 9.3^{+2.0}_{-1.5}$ keV.

\subsection{X-ray spatial analysis and Mass determination}

The distribution of the soft X-ray emission is clearly elongated in
the NE direction, resembling the distribution of cluster galaxies in
the core (see Fig.~\ref{f:xfield} and also
\citeauthor{2008A&A...489..981L} \citeyear{2008A&A...489..981L}).
 Inspection of the surface brightness (SB) profile readily shows the
presence of excess emission in the central $\sim\! 40$ kpc,
which indicates the presence of a cool core, a clear signature of the
dynamically relaxed state of the cluster (see
\cite{2008A&A...483...35S} for a discussion of cool cores in this as
well as in other high-$z$ clusters).  Here, we report on the modeling of
the SB profile, enabled by the angular resolution and depth of the
\Chandra observations, which is then used to estimate the mass of
XMM2235.

\begin{figure}
  \centering
  \includegraphics[width=\columnwidth,angle=0]{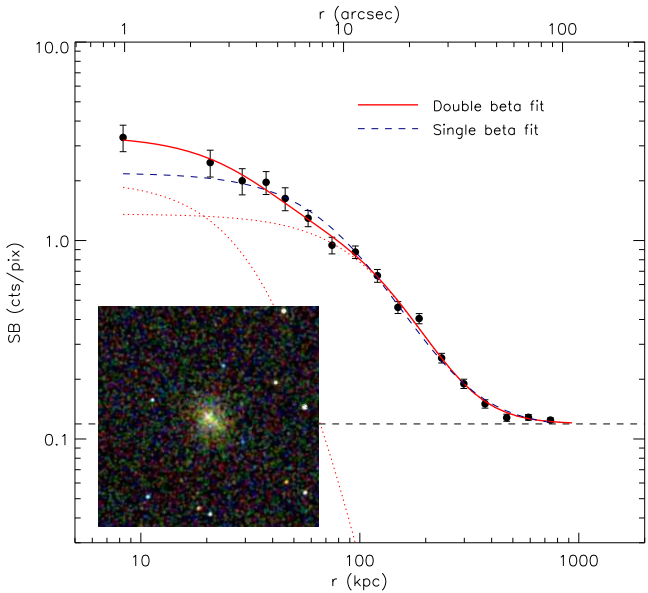}
  \caption{Surface brightness profile from the \Chandra soft X-ray
    emission of XMM2235 and best fit beta models. The dotted curves
    indicate the two components of double-$\beta$ model (solid line)
    as described in the text. Note the excess of emission in the
      central $\sim\! 40$ kpc region respect to a single-$\beta$
      model. The inset shows an X-ray color image (bluer sources
    have harder spectra) of the 2 Mpc region over which the fit is
    performed.}
\label{f:SBfit}
\end{figure}

The presence of a cool core is generally revealed by a gas density
excess in the central regions of clusters, compared to a single
$\beta-$model profile, so as to guarantee hydrostatic equilibrium.  This
is clearly illustrated in Fig.\ref{f:SBfit}. By using a projected
single $\beta-$model plus background, $\Sigma(r, R_c,
\beta)=[1+(r/R_c)^2]^{-3\beta+1/2}+B$, we obtain a poor fit in the
inner region with $R_c=90.1\pm 12.8$ kpc,\, $\beta=0.61\pm0.05$ with a
reduced $\chi^2$ of 2.05. A much better fit is obtained with a double
$\beta-$model fit,
$\Sigma(r,R_{c1},\beta_1)+\Sigma(r,R_{c2},\beta_2)+B$, which yields
$R_{c1}=45.6\pm 11.5$ kpc, $\beta_1=1.0$ (fixed), $R_{c2}=159.2\pm
11.2$ kpc, $\beta_2=0.8\pm 0.05$, $\chi_r^2=1.03$.

This information, assuming hydrostatic equilibrium and isothermality
of the gas, leads to a deprojected gas density profile and to an estimate of
the total mass. We defer to a forthcoming paper (Rosati et al. in
prep.) a detailed discussion of the X-ray mass profile of XMM2235,
with a full deprojection analysis down to the core and a measurement
of the gas fraction, as well as independent measurements of the total
mass at different radii from a dynamical analysis of all cluster
members and the modeling of a strong lensing system.
We note here that the total gravitating mass profile at large radii
($r\gg r_c$) is very well approximated by a single $\beta-$ model for
the gas density profile which leads to the well-known expression:
$M_{\rm tot}(<r) = \frac{3 \ \beta \ T_{\rm gas} \ r_{\rm c}}{G \mu
  m_{\rm p}} \frac{x^3}{ 1+x^2 }$, where $x=r/r_{\rm c}$, $\mu$ is the
mean molecular weight in atomic mass unit ($=0.59$), $G$ is the
gravitational constant, and $m_{\rm p}$ is the proton mass. Hence, $M_{\rm
  tot}(<r) = 1.13\times 10^{14} M_\odot \beta \ T\rm{(kev)} \
r\rm{(Mpc)}\frac{x^2}{1+x^2}$. Using the best fit temperature and the
single $\beta$-model parameters above, one obtains a total mass of
$M_{\rm tot} (<1\, h_{70}^{-1}Mpc) = (5.9\pm 1.3)\times
10^{14}\,M_\odot$.  Cluster masses are conveniently measured within
$R_{500}$, the radius of the sphere within which the cluster
overdensity is 500 times the critical density of the Universe at the
cluster redshift. Using the scaling relation among $R_{500}$,
temperature and redshift derived in \cite{2004A&A...417...13E} one
obtains for XMM2235 $R_{500}\approx 0.75\, h_{70}^{-1}$ Mpc and
therefore $M_{500}= M(<R_{500})= (4.4\pm 1.0)\times 10^{14}\,
M_\odot$.  To estimate the cluster virial mass, we have to extrapolate
our mass measurement to larger radii assuming for example a
\citet*{1996ApJ...462..563N} profile with concentration $c = 5$. This
yields $M_{\rm vir}\approx M_{200} = 1.4 M_{500}\approx 6\times
10^{14}\, M_\odot$.  Note that assuming a lower concentration,
  e.g. $c=3$, the virial mass would only be a few \% lower.  These
values for the total mass indicate that XMM2235 is the most massive
cluster at $z>1$ discovered to date (see
\citeauthor{2004A&A...417...13E} \citeyear{2004A&A...417...13E} for a
compilation of the masses of distant clusters derived from \Chandra data).

As discussed in \cite{2009ApJ...704..672J}, the X-ray mass profile,
which is based on the assumptions of hydrostatic equilibrium and an
isothermal gas, is found to be in very good agreement with the mass
density profile derived from the weak lensing analysis of XMM2235 from
deep HST/ACS data, which reveals a significant shear well beyond the
X-ray SB limit of 1\arcmin\ (or 500 kpc).  Specifically, using the
projection formula for the $\beta$-profile in
\cite{2005ApJ...634..813J}, a projected X-ray mass within 1 Mpc of
$M_{\rm tot, proj} (<1\, h_{70}^{-1}Mpc) = (9.3\pm 2.1)\times
10^{14}\,M_\odot$ is found, in excellent agreement with the one
derived from the shear lensing map.  We refer to the
\cite{2009ApJ...704..672J} paper for a discussion on how rare the
observation of such a massive cluster is, based on the survey volume
from which the cluster was discovered and the cluster mass function in
$\Lambda$CDM cosmology.

\section{Discussion and Conclusions}
\label{sec:results}

We presented a combination of spectro-photometric data from VLT and
HST, as well as deep \Chandra observations of the X-ray selected cluster XMMU
J2235.3$-$2557 at $z=1.39$ and used them to characterize the
galaxy populations of passive members, the thermodynamical
properties of the hot gas and the total mass of the system.

Despite surface brightness dimming, due to the high redshift of the
cluster, the \Chandra data show extended X-ray emission out to $\sim\!
0.5$ Mpc. The X-ray emission has a regular morphology and is clearly
elongated in the same way as the distribution of red passive galaxies
in the core as well as the major axis of the BCG. An excess of
emission in the inner 50 kpc is clearly detected and naturally
interpreted as a cool core. The spectral analysis of the \Chandra data
reveals that XMM2235 has a global temperature of $kT =
8.6_{-1.2}^{+1.3}$ keV (68\% confidence), which we find robust against
several systematics involved in the X-ray spectral analysis. If we
assume hydrostatic equilibrium, a well justified condition given the
relaxed appearance and the canonical $L_X/T_X$ ratio of the cluster,
and an isothermal gas distribution, the X-ray surface brightness
profile yields a total mass at large radii ($r\gg 100$ kpc) of $M_{\rm
  tot}(<r)\approx 5.9\times 10^{14}\, r({\rm Mpc})\, M_\odot$.

Overall, our analysis implies that XMM2235 is the hottest and most
massive bona--fide cluster discovered to date at $z>1$. These findings
are corroborated by the weak lensing analysis from deep HST/ACS data
\citep{2009ApJ...704..672J} which provides a mass profile in very good
agreement with the X-ray measurement.  The presence of a significant
cool core is additional evidence of the advanced dynamical state of the cluster
\citep[e.g.][]{1994ARA&A..32..277F}.

In characterizing the passive galaxy population of XMM2235, we have
extended the analysis of the ACS Intermediate Redshift Cluster Survey
at $0.8<z<1.3$ \citep[][]{2009ApJ...690...42M, 2007ApJ...663..164D} to
higher redshifts.  Using a large sample of spectroscopically
identified cluster members, we inferred the SFHs of passive galaxies
in the core ($R_{CL}<100$ kpc) and in the outskirts by modeling their
spectro-photometric data with spectral synthesis models. We find a
clear contrast in the underlying stellar populations of the two
samples. Core galaxies, all with photometric stellar masses in excess
of $10^{11} M_\odot$, appear to have formed at an earlier epoch with a
relatively short star formation phase ($z_F=6-4$), whereas passive
galaxies outside the core show spectral signatures that suggest a
star formation phase that is prolonged to later epochs, to redshifts as
low as $z\approx 2$. The latter are presumably the infalling
population of galaxies that are in the process of populating the red
sequence.  These results are also consistent with their mean color and
scatter of the $J-K_S$ red sequence \citep{2008A&A...489..981L}.

The on-going star formation is confined to the outskirts of XMM2235,
as galaxy members with a detectable [OII] emission line avoid the
inner 250 kpc region. Also in this respect, XMM2235 has already
reached an advanced evolutionary stage, as star formation is
suppressed in galaxies well before they reach the center, similarly to
low redshift massive clusters.

We note that the environmental age gradient we find in XMM2235, as
well as in other $z\gtrsim 1$ clusters
\citep[e.g.][]{2009ApJ...690...42M}, is significantly steeper than the
relatively mild difference found between {\sl field} and {\sl cluster}
early-type galaxies of similar stellar masses
\citep{2008A&A...488..853G}. This shows the importance of
environmental effects in driving galaxy evolution, in keeping with
expectations from hierarchical galaxy formation models
\citep[e.g.][]{2008ApJ...685..863M}.  It would be interesting to study
whether the radial age gradient of galaxy populations becomes larger
with increasing redshifts. This would require homogeneous galaxy
selection criteria in large samples of distant clusters with
homogeneous photometric measurements, a task which might be
challenging with current data.

At increasing redshifts, the observational effects of different
evolutionary histories of galaxies from cluster to cluster should
become apparent. Mild differences have been detected so far in the
most distant X-ray luminous clusters \citep[e.g.][in the mean color of
the color sequence]{2009ApJ...690...42M} of otherwise very homogeneous
populations.  \cite{2009ApJ...697..436H} carried out a detailed
photometric and morphological study of XMMXCS J2215.9-1738 at
$z=1.46$, currently the most distant spectroscopically confirmed X-ray
luminous cluster. Based on its X-ray temperature and luminosity, as
well its velocity dispersion, this cluster is less massive than
XMM2235 and appears to be in a younger dynamical state based on its
velocity distribution \citep{2007ApJ...670.1000H}. Another striking
difference between the two clusters is the lack of a dominant BCG in
XMM2215. However, as in XMM2235, the galaxy population of XMM2215 is
dominated by early-type galaxies with ages of $\sim\! 3$ Gyr, based on
the scatter of its red sequence, implying a major episode of star
formation at epochs $z_F\approx 3-5$.  Unfortunately, it is difficult
to directly compare these results with those we obtained for XMM2235
as the mean ages of the early-type galaxies were estimated with
different methods. A similar comparative study with optically selected
clusters would be very interesting but is still lacking. In general,
the study of cluster-to-cluster variance in distant clusters, leading
to different evolutionary histories of cluster galaxies, could provide
further tests for galaxy evolution models. A robust comparison among
different clusters however would require a homogeneous observing
campaign with the same set of instruments and filters so as to minimize
systematics in photometric measurements and model dependent
K-corrections.

Another notable result of our study is the measured metal abundance $Z
= 0.26^{+0.20}_{-0.16} \, Z_\odot$ of the intracluster medium (ICM)
from the detection of the Fe-K line in the \Chandra spectrum,
essentially coming from the cluster core ($R_{CL}\lesssim 200$ kpc
where most of the \Chandra counts are contained). This extends to even
larger look-back times the evidence that the ICM was already
significantly enriched in distant clusters, as found in a systematic
study of clusters at $z<1.3$ 
\citep[][]{2007A&A...462..429B, 2008ApJS..174..117M}, 
and therefore that metal enrichment of the ICM was mostly complete at early
epochs.  In light of the derived SFH of the core red galaxies this is
not unexpected. Chemical evolution models of elliptical galaxies
\citep[e.g.][]{2004MNRAS.347..968P} predict that chemical enrichment
of massive elliptical galaxies, which experience a short burst of star
formation at early epochs, occurs on a time scale of 1 Gyr. Galactic
winds distribute metals in the cluster core region on a crossing time
scale ($\lesssim 1$ Gyr). As a result, one would expect the metal
enrichment of the ICM to be completed by $z\gtrsim 2.5$ (2 Gyr earlier
than the cluster look-back time of 9 Gyr), consistently with the
derived SFH of the core galaxies.

It remains somewhat surprising that such a massive cluster, which was
discovered from a relatively small survey area (11 deg$^2$), is found
with a baryonic content in such an advanced evolutionary stage at an
epoch corresponding to 1/3 of the current age of the Universe, both in
terms of its galaxy population and of the hot intracluster gas.
Further results on the galaxy populations of XMM2235 as well as its
mass distribution, taking advantage of an augmented multi-wavelength
data set, will be presented in forthcoming papers.

Given its large mass, XMM2235 will also likely be an ideal target for
a new array of Sunayev-Zeldovich (SZ) experiments which are already
operational \citep[e.g.][]{2009ApJ...701...32S} or are becoming
available. These observations will add independent, important
information on the physics of the ICM and the total mass of XMM2235
and will be useful to calibrate the expectations of a new era of SZ
cluster surveys.

\begin{acknowledgements}
  We acknowledge the excellent support provided by the VLT staff at
  the Paranal observatory. We thank S.Ettori and S.Borgani for useful
  discussions.  PT acknowledges financial support from contract
  ASI-INAF I/088/06/0 and from the PD51 INFN grant. JSS acknowledges
  support by the Deutsche Forschungsgemeinschaft under contract
  BO702/16-2.  RG acknowledges partial support by the DFG cluster of
  excellence “Origin and Structure of the Universe”
  (www.universe-cluster.de). GL was supported by the DLR under
  contract numbers 50OX0201 and 50QR0802.
\end{acknowledgements}

\bibliographystyle{aa}
\bibliography{13099}

\begin{thebibliography}{55}
\expandafter\ifx\csname natexlab\endcsname\relax\def\natexlab#1{#1}\fi

\bibitem[{{Abraham} {et~al.}(2004){Abraham}, {Glazebrook}, {McCarthy},
  {Crampton}, {Murowinski}, {J{\o}rgensen}, {Roth}, {Hook}, {Savaglio}, {Chen},
  {Marzke}, \& {Carlberg}}]{2004AJ....127.2455A}
{Abraham}, R.~G., {Glazebrook}, K., {McCarthy}, P.~J., {et~al.} 2004, \aj, 127,
  2455

\bibitem[{{Anders} \& {Grevesse}(1989)}]{1989GeCoA..53..197A}
{Anders}, E. \& {Grevesse}, N. 1989, \gca, 53, 197

\bibitem[{{Arnaud}(1996)}]{1996ASPC..101...17A}
{Arnaud}, K.~A. 1996, in Astronomical Society of the Pacific Conference Series,
  Vol. 101, Astronomical Data Analysis Software and Systems V, ed. G.~H.
  {Jacoby} \& J.~{Barnes}, 17

\bibitem[{{Asplund} {et~al.}(2005){Asplund}, {Grevesse}, \&
  {Sauval}}]{2005ASPC..336...25A}
{Asplund}, M., {Grevesse}, N., \& {Sauval}, A.~J. 2005, in Astronomical Society
  of the Pacific Conference Series, Vol. 336, Cosmic Abundances as Records of
  Stellar Evolution and Nucleosynthesis, ed. T.~G. {Barnes}, III \& F.~N.
  {Bash}, 25

\bibitem[{{Balestra} {et~al.}(2007){Balestra}, {Tozzi}, {Ettori}, {Rosati},
  {Borgani}, {Mainieri}, {Norman}, \& {Viola}}]{2007A&A...462..429B}
{Balestra}, I., {Tozzi}, P., {Ettori}, S., {et~al.} 2007, \aap, 462, 429

\bibitem[{{Beers} {et~al.}(1990){Beers}, {Flynn}, \&
  {Gebhardt}}]{1990AJ....100...32B}
{Beers}, T.~C., {Flynn}, K., \& {Gebhardt}, K. 1990, \aj, 100, 32

\bibitem[{{Bignamini} {et~al.}(2008){Bignamini}, {Tozzi}, {Borgani}, {Ettori},
  \& {Rosati}}]{2008A&A...489..967B}
{Bignamini}, A., {Tozzi}, P., {Borgani}, S., {Ettori}, S., \& {Rosati}, P.
  2008, \aap, 489, 967

\bibitem[{{Blakeslee} {et~al.}(2003){Blakeslee}, {Anderson}, {Meurer},
  {Ben{\'{\i}}tez}, \& {Magee}}]{2003ASPC..295..257B}
{Blakeslee}, J.~P., {Anderson}, K.~R., {Meurer}, G.~R., {Ben{\'{\i}}tez}, N.,
  \& {Magee}, D. 2003, in Astronomical Society of the Pacific Conference
  Series, Vol. 295, Astronomical Data Analysis Software and Systems XII, ed.
  H.~E. {Payne}, R.~I. {Jedrzejewski}, \& R.~N. {Hook}, 257

\bibitem[{{Boehringer} {et~al.}(2005){Boehringer}, {Mullis}, {Rosati}, {Lamer},
  {Fassbender}, {Schwope}, \& {Schuecker}}]{2005Msngr.120...33B}
{Boehringer}, H., {Mullis}, C., {Rosati}, P., {et~al.} 2005, The Messenger,
  120, 33

\bibitem[{{Bruzual} \& {Charlot}(2003)}]{2003MNRAS.344.1000B}
{Bruzual}, G. \& {Charlot}, S. 2003, \mnras, 344, 1000

\bibitem[{{Cimatti} {et~al.}(2008){Cimatti}, {Cassata}, {Pozzetti}, {Kurk},
  {Mignoli}, {Renzini}, {Daddi}, {Bolzonella}, {Brusa}, {Rodighiero},
  {Dickinson}, {Franceschini}, {Zamorani}, {Berta}, {Rosati}, \&
  {Halliday}}]{2008A&A...482...21C}
{Cimatti}, A., {Cassata}, P., {Pozzetti}, L., {et~al.} 2008, \aap, 482, 21

\bibitem[{{Demarco} {et~al.}(2007){Demarco}, {Rosati}, {Lidman}, {Girardi},
  {Nonino}, {Rettura}, {Strazzullo}, {van der Wel}, {Ford}, {Mainieri},
  {Holden}, {Stanford}, {Blakeslee}, {Gobat}, {Postman}, {Tozzi}, {Overzier},
  {Zirm}, {Ben{\'{\i}}tez}, {Homeier}, {Illingworth}, {Infante}, {Jee}, {Mei},
  {Menanteau}, {Motta}, {Zheng}, {Clampin}, \& {Hartig}}]{2007ApJ...663..164D}
{Demarco}, R., {Rosati}, P., {Lidman}, C., {et~al.} 2007, \apj, 663, 164

\bibitem[{{Dickey} \& {Lockman}(1990)}]{1990ARA&A..28..215D}
{Dickey}, J.~M. \& {Lockman}, F.~J. 1990, \araa, 28, 215

\bibitem[{{Eisenhardt} {et~al.}(2008){Eisenhardt}, {Brodwin}, {Gonzalez},
  {Stanford}, {Stern}, {Barmby}, {Brown}, {Dawson}, {Dey}, {Doi}, {Galametz},
  {Jannuzi}, {Kochanek}, {Meyers}, {Morokuma}, \&
  {Moustakas}}]{2008ApJ...684..905E}
{Eisenhardt}, P.~R.~M., {Brodwin}, M., {Gonzalez}, A.~H., {et~al.} 2008, \apj,
  684, 905

\bibitem[{{Ettori} {et~al.}(2004){Ettori}, {Tozzi}, {Borgani}, \&
  {Rosati}}]{2004A&A...417...13E}
{Ettori}, S., {Tozzi}, P., {Borgani}, S., \& {Rosati}, P. 2004, \aap, 417, 13

\bibitem[{{Fabian}(1994)}]{1994ARA&A..32..277F}
{Fabian}, A.~C. 1994, \araa, 32, 277

\bibitem[{{Fassbender}(2007)}]{2008arXiv0806.0861F}
{Fassbender}, R. 2007, PhD thesis, Ludwig-Maximilians-Universitaet Muenchen,
  arXiv:0806.0861

\bibitem[{{Ford} {et~al.}(2004){Ford}, {Postman}, {Blakeslee}, {Demarco},
  {Jee}, {Rosati}, {Holden}, {Homeier}, {Illingworth}, \&
  {White}}]{2004ASSL..319..459F}
{Ford}, H., {Postman}, M., {Blakeslee}, J.~P., {et~al.} 2004, in Astrophysics
  and Space Science Library, Vol. 319, Penetrating Bars Through Masks of Cosmic
  Dust, ed. D.~L. {Block}, I.~{Puerari}, K.~C. {Freeman}, R.~{Groess}, \& E.~K.
  {Block}, 459, arXiv:0408165

\bibitem[{{Gobat} {et~al.}(2008){Gobat}, {Rosati}, {Strazzullo}, {Rettura},
  {Demarco}, \& {Nonino}}]{2008A&A...488..853G}
{Gobat}, R., {Rosati}, P., {Strazzullo}, V., {et~al.} 2008, \aap, 488, 853

\bibitem[{{Grevesse} \& {Sauval}(1998)}]{1998SSRv...85..161G}
{Grevesse}, N. \& {Sauval}, A.~J. 1998, Space Science Reviews, 85, 161

\bibitem[{{Hilton} {et~al.}(2007){Hilton}, {Collins}, {Stanford}, {Lidman},
  {Dawson}, {Davidson}, {Kay}, {Liddle}, {Mann}, {Miller}, {Nichol}, {Romer},
  {Sabirli}, {Viana}, \& {West}}]{2007ApJ...670.1000H}
{Hilton}, M., {Collins}, C.~A., {Stanford}, S.~A., {et~al.} 2007, \apj, 670,
  1000

\bibitem[{{Hilton} {et~al.}(2009){Hilton}, {Stanford}, {Stott}, {Collins},
  {Hoyle}, {Davidson}, {Hosmer}, {Kay}, {Liddle}, {Lloyd-Davies}, {Mann},
  {Mehrtens}, {Miller}, {Nichol}, {Romer}, {Sabirli}, {Sahl{\'e}n}, {Viana},
  {West}, {Barbary}, {Dawson}, {Meyers}, {Perlmutter}, {Rubin}, \&
  {Suzuki}}]{2009ApJ...697..436H}
{Hilton}, M., {Stanford}, S.~A., {Stott}, J.~P., {et~al.} 2009, \apj, 697, 436

\bibitem[{{Holden} {et~al.}(2007){Holden}, {Illingworth}, {Franx}, {Blakeslee},
  {Postman}, {Kelson}, {van der Wel}, {Demarco}, {Magee}, {Tran}, {Zirm},
  {Ford}, {Rosati}, \& {Homeier}}]{2007ApJ...670..190H}
{Holden}, B.~P., {Illingworth}, G.~D., {Franx}, M., {et~al.} 2007, \apj, 670,
  190

\bibitem[{{Jee} {et~al.}(2009){Jee}, {Rosati}, {Ford}, {Dawson}, {Lidman},
  {Perlmutter}, {Demarco}, {Strazzullo}, {Mullis}, {B{\"o}hringer}, \&
  {Fassbender}}]{2009ApJ...704..672J}
{Jee}, M.~J., {Rosati}, P., {Ford}, H.~C., {et~al.} 2009, \apj, 704, 672

\bibitem[{{Jee} {et~al.}(2005){Jee}, {White}, {Ford}, {Blakeslee},
  {Illingworth}, {Coe}, \& {Tran}}]{2005ApJ...634..813J}
{Jee}, M.~J., {White}, R.~L., {Ford}, H.~C., {et~al.} 2005, \apj, 634, 813

\bibitem[{{Kodama} {et~al.}(2007){Kodama}, {Tanaka}, {Kajisawa}, {Kurk},
  {Venemans}, {De Breuck}, {Vernet}, \& {Lidman}}]{2007MNRAS.377.1717K}
{Kodama}, T., {Tanaka}, I., {Kajisawa}, M., {et~al.} 2007, \mnras, 377, 1717

\bibitem[{{Kriek} {et~al.}(2006){Kriek}, {van Dokkum}, {Franx}, {Quadri},
  {Gawiser}, {Herrera}, {Illingworth}, {Labb{\'e}}, {Lira}, {Marchesini},
  {Rix}, {Rudnick}, {Taylor}, {Toft}, {Urry}, \& {Wuyts}}]{2006ApJ...649L..71K}
{Kriek}, M., {van Dokkum}, P.~G., {Franx}, M., {et~al.} 2006, \apjl, 649, L71

\bibitem[{{Lamer} {et~al.}(2008){Lamer}, {Hoeft}, {Kohnert}, {Schwope}, \&
  {Storm}}]{2008A&A...487L..33L}
{Lamer}, G., {Hoeft}, M., {Kohnert}, J., {Schwope}, A., \& {Storm}, J. 2008,
  \aap, 487, L33

\bibitem[{{Lidman} {et~al.}(2008){Lidman}, {Rosati}, {Tanaka}, {Strazzullo},
  {Demarco}, {Mullis}, {Ageorges}, {Kissler-Patig}, {Petr-Gotzens}, \&
  {Selman}}]{2008A&A...489..981L}
{Lidman}, C., {Rosati}, P., {Tanaka}, M., {et~al.} 2008, \aap, 489, 981

\bibitem[{{Liedahl} {et~al.}(1995){Liedahl}, {Osterheld}, \&
  {Goldstein}}]{1995ApJ...438L.115L}
{Liedahl}, D.~A., {Osterheld}, A.~L., \& {Goldstein}, W.~H. 1995, \apjl, 438,
  L115

\bibitem[{{Maughan} {et~al.}(2008){Maughan}, {Jones}, {Forman}, \& {Van
  Speybroeck}}]{2008ApJS..174..117M}
{Maughan}, B.~J., {Jones}, C., {Forman}, W., \& {Van Speybroeck}, L. 2008,
  \apjs, 174, 117

\bibitem[{{Mei} {et~al.}(2009){Mei}, {Holden}, {Blakeslee}, {Ford}, {Franx},
  {Homeier}, {Illingworth}, {Jee}, {Overzier}, {Postman}, {Rosati}, {Van der
  Wel}, \& {Bartlett}}]{2009ApJ...690...42M}
{Mei}, S., {Holden}, B.~P., {Blakeslee}, J.~P., {et~al.} 2009, \apj, 690, 42

\bibitem[{{Mei} {et~al.}(2006){Mei}, {Holden}, {Blakeslee}, {Rosati},
  {Postman}, {Jee}, {Rettura}, {Sirianni}, {Demarco}, {Ford}, {Franx},
  {Homeier}, \& {Illingworth}}]{2006ApJ...644..759M}
{Mei}, S., {Holden}, B.~P., {Blakeslee}, J.~P., {et~al.} 2006, \apj, 644, 759

\bibitem[{{Menci} {et~al.}(2008){Menci}, {Rosati}, {Gobat}, {Strazzullo},
  {Rettura}, {Mei}, \& {Demarco}}]{2008ApJ...685..863M}
{Menci}, N., {Rosati}, P., {Gobat}, R., {et~al.} 2008, \apj, 685, 863

\bibitem[{{Mullis} {et~al.}(2005){Mullis}, {Rosati}, {Lamer}, {B{\"o}hringer},
  {Schwope}, {Schuecker}, \& {Fassbender}}]{2005ApJ...623L..85M}
{Mullis}, C.~R., {Rosati}, P., {Lamer}, G., {et~al.} 2005, \apjl, 623, L85

\bibitem[{{Navarro} {et~al.}(1996){Navarro}, {Frenk}, \&
  {White}}]{1996ApJ...462..563N}
{Navarro}, J.~F., {Frenk}, C.~S., \& {White}, S.~D.~M. 1996, \apj, 462, 563

\bibitem[{{Persson} {et~al.}(1998){Persson}, {Murphy}, {Krzeminski}, {Roth}, \&
  {Rieke}}]{1998AJ....116.2475P}
{Persson}, S.~E., {Murphy}, D.~C., {Krzeminski}, W., {Roth}, M., \& {Rieke},
  M.~J. 1998, \aj, 116, 2475

\bibitem[{{Pickles}(1998)}]{1998PASP..110..863P}
{Pickles}, A.~J. 1998, \pasp, 110, 863

\bibitem[{{Pipino} \& {Matteucci}(2004)}]{2004MNRAS.347..968P}
{Pipino}, A. \& {Matteucci}, F. 2004, \mnras, 347, 968

\bibitem[{{Postman} {et~al.}(2005){Postman}, {Franx}, {Cross}, {Holden},
  {Ford}, {Illingworth}, {Goto}, {Demarco}, {Rosati}, {Blakeslee}, {Tran},
  {Ben{\'{\i}}tez}, {Clampin}, {Hartig}, {Homeier}, {Ardila}, {Bartko},
  {Bouwens}, {Bradley}, {Broadhurst}, {Brown}, {Burrows}, {Cheng}, {Feldman},
  {Golimowski}, {Gronwall}, {Infante}, {Kimble}, {Krist}, {Lesser}, {Martel},
  {Mei}, {Menanteau}, {Meurer}, {Miley}, {Motta}, {Sirianni}, {Sparks}, {Tran},
  {Tsvetanov}, {White}, \& {Zheng}}]{2005ApJ...623..721P}
{Postman}, M., {Franx}, M., {Cross}, N.~J.~G., {et~al.} 2005, \apj, 623, 721

\bibitem[{{Renzini}(2006)}]{2006ARA&A..44..141R}
{Renzini}, A. 2006, \araa, 44, 141

\bibitem[{{Rettura} {et~al.}(2008){Rettura}, {Rosati}, {Nonino}, {Fosbury},
  {Gobat}, {Menci}, {Strazzullo}, {Mei}, {Demarco}, \&
  {Ford}}]{2008arXiv0806.4604R}
{Rettura}, A., {Rosati}, P., {Nonino}, M., {et~al.} 2008, arXiv:0806.4604, \apj, in press

\bibitem[{{Rosati} {et~al.}(2002){Rosati}, {Borgani}, \&
  {Norman}}]{2002ARA&A..40..539R}
{Rosati}, P., {Borgani}, S., \& {Norman}, C. 2002, \araa, 40, 539

\bibitem[{{Rosati} {et~al.}(2004){Rosati}, {Tozzi}, {Ettori}, {Mainieri},
  {Demarco}, {Stanford}, {Lidman}, {Nonino}, {Borgani}, {Della Ceca},
  {Eisenhardt}, {Holden}, \& {Norman}}]{2004AJ....127..230R}
{Rosati}, P., {Tozzi}, P., {Ettori}, S., {et~al.} 2004, \aj, 127, 230

\bibitem[{{Santos} {et~al.}(2009){Santos}, {Rosati}, {Gobat}, {Lidman},
  {Dawson}, {Perlmutter}, {B{\"o}hringer}, {Balestra}, {Mullis}, {Fassbender},
  {Kohnert}, {Lamer}, {Rettura}, {Rit{\'e}}, \&
  {Schwope}}]{2009A&A...501...49S}
{Santos}, J.~S., {Rosati}, P., {Gobat}, R., {et~al.} 2009, \aap, 501, 49

\bibitem[{{Santos} {et~al.}(2008){Santos}, {Rosati}, {Tozzi}, {B{\"o}hringer},
  {Ettori}, \& {Bignamini}}]{2008A&A...483...35S}
{Santos}, J.~S., {Rosati}, P., {Tozzi}, P., {et~al.} 2008, \aap, 483, 35

\bibitem[{{Sirianni} {et~al.}(2005){Sirianni}, {Jee}, {Ben{\'{\i}}tez},
  {Blakeslee}, {Martel}, {Meurer}, {Clampin}, {De Marchi}, {Ford}, {Gilliland},
  {Hartig}, {Illingworth}, {Mack}, \& {McCann}}]{2005PASP..117.1049S}
{Sirianni}, M., {Jee}, M.~J., {Ben{\'{\i}}tez}, N., {et~al.} 2005, \pasp, 117,
  1049

\bibitem[{{Stanford} {et~al.}(2001){Stanford}, {Holden}, {Rosati}, {Tozzi},
  {Borgani}, {Eisenhardt}, \& {Spinrad}}]{2001ApJ...552..504S}
{Stanford}, S.~A., {Holden}, B., {Rosati}, P., {et~al.} 2001, \apj, 552, 504

\bibitem[{{Stanford} {et~al.}(2006){Stanford}, {Romer}, {Sabirli}, {Davidson},
  {Hilton}, {Viana}, {Collins}, {Kay}, {Liddle}, {Mann}, {Miller}, {Nichol},
  {West}, {Conselice}, {Spinrad}, {Stern}, \& {Bundy}}]{2006ApJ...646L..13S}
{Stanford}, S.~A., {Romer}, A.~K., {Sabirli}, K., {et~al.} 2006, \apjl, 646,
  L13

\bibitem[{{Staniszewski} {et~al.}(2009){Staniszewski}, {Ade}, {Aird}, {Benson},
  {Bleem}, {Carlstrom}, {Chang}, {Cho}, {Crawford}, {Crites}, {de Haan},
  {Dobbs}, {Halverson}, {Holder}, {Holzapfel}, {Hrubes}, {Joy}, {Keisler},
  {Lanting}, {Lee}, {Leitch}, {Loehr}, {Lueker}, {McMahon}, {Mehl}, {Meyer},
  {Mohr}, {Montroy}, {Ngeow}, {Padin}, {Plagge}, {Pryke}, {Reichardt}, {Ruhl},
  {Schaffer}, {Shaw}, {Shirokoff}, {Spieler}, {Stalder}, {Stark},
  {Vanderlinde}, {Vieira}, {Zahn}, \& {Zenteno}}]{2009ApJ...701...32S}
{Staniszewski}, Z., {Ade}, P.~A.~R., {Aird}, K.~A., {et~al.} 2009, \apj, 701,
  32

\bibitem[{{van Dokkum} \& {van der Marel}(2007)}]{2007ApJ...655...30V}
{van Dokkum}, P.~G. \& {van der Marel}, R.~P. 2007, \apj, 655, 30

\bibitem[{{Voit}(2005)}]{2005RvMP...77..207V}
{Voit}, G.~M. 2005, Reviews of Modern Physics, 77, 207

\bibitem[{{Wilms} {et~al.}(2000){Wilms}, {Allen}, \&
  {McCray}}]{2000ApJ...542..914W}
{Wilms}, J., {Allen}, A., \& {McCray}, R. 2000, \apj, 542, 914

\bibitem[{{Wilson} {et~al.}(2009){Wilson}, {Muzzin}, {Yee}, {Lacy}, {Surace},
  {Gilbank}, {Blindert}, {Hoekstra}, {Majumdar}, {Demarco}, {Gardner},
  {Gladders}, \& {Lonsdale}}]{2009ApJ...698.1943W}
{Wilson}, G., {Muzzin}, A., {Yee}, H.~K.~C., {et~al.} 2009, \apj, 698, 1943

\bibitem[{{Zirm} {et~al.}(2008){Zirm}, {Stanford}, {Postman}, {Overzier},
  {Blakeslee}, {Rosati}, {Kurk}, {Pentericci}, {Venemans}, {Miley},
  {R{\"o}ttgering}, {Franx}, {van der Wel}, {Demarco}, \& {van
  Breugel}}]{2008ApJ...680..224Z}
{Zirm}, A.~W., {Stanford}, S.~A., {Postman}, M., {et~al.} 2008, \apj, 680, 224

\end{thebibliography}

\end{document}